\documentclass[sn-mathphys,Numbered,twocolumn]{sn-jnl}
\usepackage{graphicx}%
\usepackage{multirow}%
\usepackage{amsmath,amssymb,amsfonts}%
\usepackage{amsthm}%
\usepackage{mathrsfs}%
\usepackage[title]{appendix}%
\usepackage{xcolor}%
\usepackage{textcomp}%
\usepackage{manyfoot}%
\usepackage{booktabs}%
\usepackage{algorithm}%
\usepackage{algorithmicx}%
\usepackage{algpseudocode}%
\usepackage{listings}%
\usepackage{caption}
\usepackage{multicol}
\usepackage{pdfcomment} 
\usepackage{datetime2}
\usepackage{comment}
\usepackage{textcomp}

\begin{document}

\newgeometry{a4paper,margin=0.5in,twocolumn}



\title[Article Title]{\textbf{The Non-collinear Path to Topological Superconductivity}}

\author*[1]{\fnm{Reiner} \sur{Br\"uning}} \email{rbruenin@physnet.uni-hamburg.de}
\author[2]{\fnm{Jasmin} \sur{Bedow}}

\author[1,3]{\fnm{Roberto} \sur{Lo Conte}}

\author*[1]{\fnm{Kirsten} \sur{von Bergmann}} \email{kirsten.von.bergmann@physik.uni-hamburg.de}
\author[2]{\fnm{Dirk K.} \sur{Morr}}
\author[1]{\fnm{Roland} \sur{Wiesendanger}}

\affil[1]{\orgdiv{Department of Physics, University of Hamburg}, \postcode{20355} \city{Hamburg}, \country{Germany}}

\affil[2]{\orgdiv{
Department of Physics, University of Illinois at Chicago},  \city{Chicago}, \postcode{IL 60646}, \country{USA}}

\affil[3]{\orgdiv{Zernike Institute for Advanced Materials, University of Groningen}, \city{9747 AG Groningen}, \country{The Netherlands}}


\abstract{

Combining spin textures in ultra-thin films with conventional superconductors has emerged as a powerful and versatile platform for designing topologically non-trivial superconducting phases as well as spin-triplet Cooper pairs. 
As a consequence, two-dimensional magnet-superconductor hybrids (2D MSHs) are promising candidate systems to realize devices for topology-based quantum technologies and superconducting spintronics. So far, studies have focused mostly on systems hosting collinear ferromagnets or antiferromagnets. However, topologically non-trivial phases have been predicted to emerge in MSH systems with non-collinear spin textures as well. 
In this article, we present the experimental discovery of topological superconductivity in the MSH system Fe/Ta(110) where a magnetic spiral is realized in the Fe monolayer on the surface of the  $s$-wave superconductor Ta. By combining low-temperature spin-polarized scanning tunneling microscopy measurements with theoretical modeling, we are able to conclude that the system is in a topological nodal-point superconducting phase with low-energy edge modes. Due to the non-collinear spin texture in our MSH system, these edge modes exhibit a magnetization direction-dependent dispersion. Furthermore, we identify direct signatures of Rashba spin-orbit coupling in the experimentally measured differential tunneling conductance. The present work realizes a non-collinear spin texture-based path to topological superconductivity.}

\maketitle

\vspace*{0.3\baselineskip}

Topological superconductivity in bulk as well as in low-dimensional hybrid structures has become a very active field of research in recent years \cite{andoAnnu.Rev.Condens.MatterPhys.2015,satoRep.Prog.Phys.2017}. Early attempts to establish topological superconductivity in two-dimensional magnet-superconductor hybrids (2D-MSH) focused primarily on ferromagnetic films in proximity to $s$-wave superconductors~\cite{MenardNAT-COMM2017,Palacio-MoralesSCI-ADV2019,KezilebiekeNAT2020}. Such a system, in combination with spin-orbit coupling (SOC), is expected to generate a topological phase with a hard superconducting gap and dispersing chiral edge modes. This topological phase can be classified by a topological invariant, the Chern number, which relates to the number of chiral edge modes~\cite{LiNAT-COMM2016}. More recently, the attempts to establish topologically non-trivial states have been extended to collinear antiferromagnetic states in two-dimensional structures on conventional superconductors~\cite{LoContePRB2022,BazarnikNAT-COMM2023,SoldiniNAT-PHYS2023,Zhang2019,KieuPRB2023}. In these systems,  new types of topological phases can emerge such as topological nodal point superconductivity~\cite{BazarnikNAT-COMM2023,Zhang2019,KieuPRB2023} and topological crystalline superconductivity~\cite{SoldiniNAT-PHYS2023}. All of the above-mentioned studies have established that 
2D-MSH systems provide
a very powerful approach for realizing novel topological phases of matter.

While so far, only collinear  ferromagnetic and antiferromagnetic spin textures -- a small sub-group of all the possible spin textures available -- have been investigated as a source of non-trivial topology, there have been numerous theoretical predictions of novel topological superconducting phases in non-collinear~\cite{NakosaiPRB2013,ChatterjeePRB2024-1,ChatterjeePRB2024-2} and non-coplanar~\cite{BedowPRB2020,MascotQ-MAT2021,SteffensenPRR2022} spin textures in proximity to a superconductor. 
However, the experimental observation of these predictions has so far been elusive.

Here, we report the experimental discovery of topological nodal-point superconductivity in a non-collinear 
2D-MSH 
system, where a single atomic layer of iron (Fe) is grown on top of a clean superconducting Ta(110) substrate. The presence of Ta, with large SOC, is crucial for the stabilization of a cycloidal spin spiral ground state in the Fe monolayer due to the Dzyaloshinskii-Moriya interaction~\cite{rozsaPhys.Rev.B2015}. Using scanning tunneling microscopy (STM) and spectroscopy, we identify the existence 
of the spin spiral, and gain insight into the 
spatially-resolved
electronic structure of the hybrid system by measuring the local differential tunneling conductance, d$I$/d$V$, using both spin-polarized and 
non-magnetic
tips. Complementary theoretical studies allow us to prove the existence of a topological nodal point superconducting (TNPSC) phase in the system, identifying its characteristic experimental signatures. In particular, we verify the existence of a topologically protected edge mode along the [001] edge of a magnetic island, which is absent along the [1$\bar{1}$0] and [1$\bar{1}$1] edges. We demonstrate that the dispersion of the edge mode and its spectral weight depends sensitively on the magnetization direction at the island edge. A comparison of these results with the experimentally measured d$I$/d$V$ signal along island edges yields good agreement, providing strong evidence for the existence of a TNPSC phase arising from a non-collinear spin texture.



\begin{figure}[!htbp]
\centering
\includegraphics[width=\columnwidth]{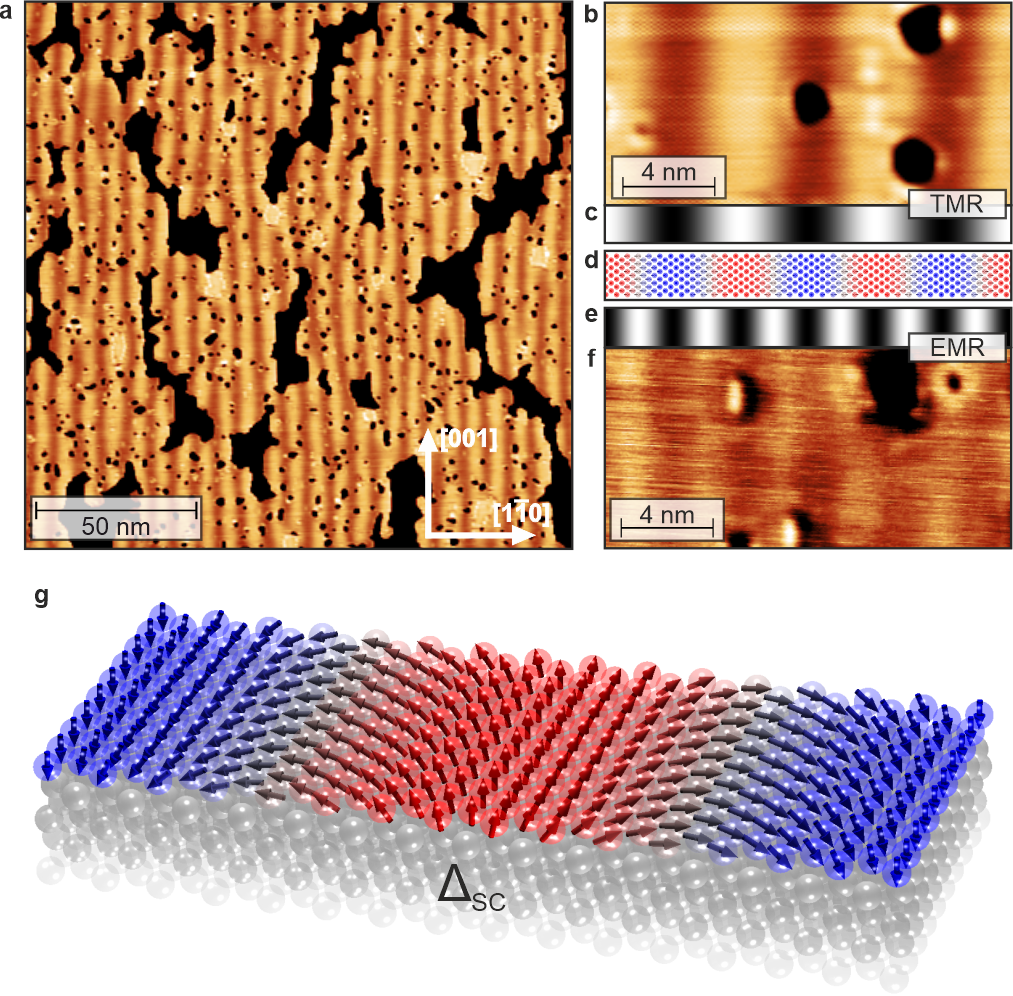}
\vspace{0.01cm}
\captionsetup{font={small}}
\caption{\textbf{Magnetic characterization of the MSH system.}
\textbf{a} SP-STM constant-current image of a sample with 0.8 atomic layers of Fe on Ta(110); the pseudomorphic Fe monolayer areas can be identified by their orange color, the stripes originate from the magnetic spin spiral state.
\textbf{b} Closer view of a SP-STM constant-current measurement of the spin spiral in the Fe monolayer.
\textbf{c} Expected SP-STM signal due to the TMR effect for a spin spiral. 
\textbf{d} Sketch of a homogeneous spin spiral; red and blue indicate up and down magnetization directions, respectively.
\textbf{e} Expected signal for a spin spiral using a non-magnetic tip due to EMR effects. 
\textbf{f} STM constant-current image obtained using a non-magnetic tip, revealing a periodic pattern with half of the spin spiral period, in agreement with EMR contrast. 
\textbf{g} Perspective view of the studied MSH system.}
\label{fig:Introduction}
\end{figure}

\noindent

An overview spin-polarized (SP-)STM image with about 80 percent coverage of Fe on Ta(110) is shown in Fig.~\ref{fig:Introduction}a. 
The Fe monolayer (orange) grows pseudomorphically, and the stripes along the $[001]$ direction are of magnetic origin, visible due to the spin-polarized tip used in this measurement~\cite{wiesendanger_spin_2009,bergmann_interface-induced_2014}. We identify a spin spiral as the magnetic ground state of the Fe monolayer propagating along $[1\overline{1}0]$ with a period of about $6$~nm ($5.2-7.0$~nm depending on the location). 
A closer view of the spin spiral is shown in Fig.~\ref{fig:Introduction}b, where bright areas are parallel to the tip magnetization direction and dark areas are antiparallel to it (or vice versa).
This imaging mechanism is based on the tunnel magnetoresistance (TMR) effect and the signal scales with the cosine of the angle enclosed by tip and sample magnetization, i.e., it directly reflects the magnetic periodicity. This is shown in Figs.~\ref{fig:Introduction}c and \ref{fig:Introduction}d, where the expected imaging contrast 
is directly compared to a sketch of the respective spin spiral.


When a 
non-magnetic
tip is used, as for the STM image displayed in Fig.~\ref{fig:Introduction}f, a stripe pattern with half the spin spiral period becomes visible, albeit with much lower intensity. 
This
has previously been observed in spin spiral systems and has been ascribed to either a SOC-induced tunnel anisotropic magnetoresistance (TAMR)~\cite{BergmannPRB2012, herve_stabilizing_2018} or -- in the case of inhomogeneous spin spirals -- to a non-collinear magnetoresistance (NCMR) originating from spin mixing~\cite{HannekenNN2015,crumNC2015,perini_electrical_2019} (see expected imaging contrast 
in Fig.~\ref{fig:Introduction}e). Because we cannot distinguish experimentally which of the two effects is dominating in our system, we ascribe this observation to an electronic magnetoresistance (EMR) effect in the following, as both effects are related to slight changes of the local electronic states due to the 
non-collinearity of the 
spin texture. A perspective view of the established 2D-MSH model system is displayed in Fig.~\ref{fig:Introduction}g.

\begin{figure}[!htbp]
\centering
\includegraphics[width=\columnwidth]{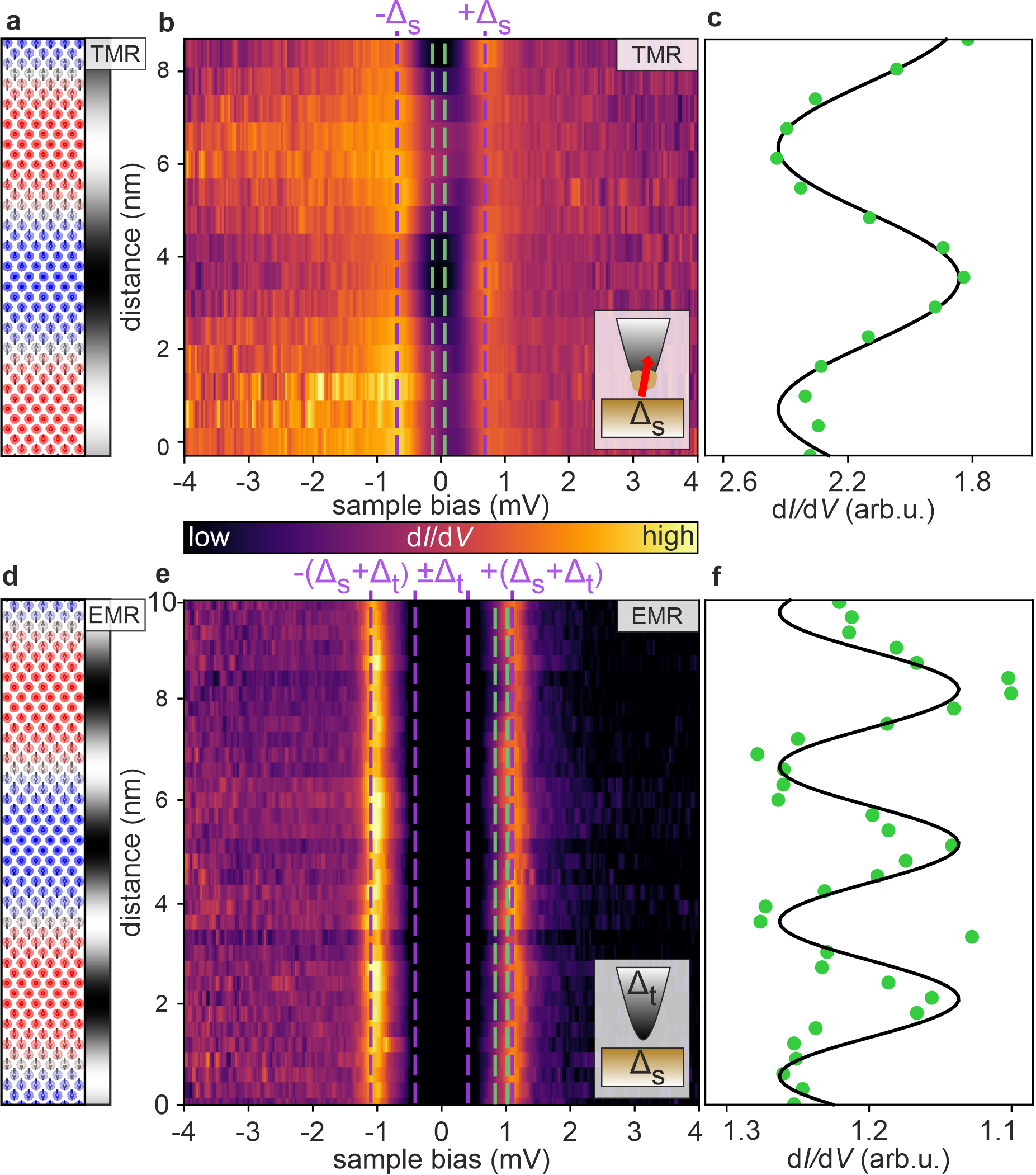}
\vspace{0.01cm}
\captionsetup{width=\columnwidth}
\small
\flushleft
\caption{
\textbf{Spectroscopic characterization of the MSH system.} \textbf{a} Spin configuration of the magnetic spin spiral and the expected imaging contrast due to the TMR effect. 
\textbf{b} Spin-resolved d$I$/d$V$ spectroscopy measurements obtained with a 
spin-polarized tip along the spin spiral propagation direction. 
\textbf{c} d$I$/d$V$ intensities along the spin spiral averaged over an energy range of $-0.13$~mV to $+0.06$~mV, as indicated by the green lines in \textbf{b}; the solid line represents a cosine function with the spin spiral period and serves as a guide to the eye.
\textbf{d} Spin configuration of the magnetic spin spiral and the expected imaging contrast due to the EMR effect. 
\textbf{e} spin-averaged d$I$/d$V$ spectroscopy measurements obtained with a superconducting tip along the spin spiral. 
\textbf{f} d$I$/d$V$ intensities along the spin spiral averaged over an energy range of $+0.80$~mV to $+1.02$~mV, as indicated by the green lines in \textbf{e}; the cosine function with half the spin spiral period serves as a guide to the eye.}
\label{fig:Periods}
\end{figure}

\noindent

A sketch of the magnetic spin texture with the expected signal due to the TMR effect is displayed again in Fig.~\ref{fig:Periods}a, while in Fig.~\ref{fig:Periods}b, we present the spin-resolved d$I$/d$V$ measured as a function of the sample bias and the position within the spin spiral.
%
These 
measurements
reveal the existence of a superconducting gap in the Fe monolayer with the coherence peaks indicated by the purple dashed lines at $\pm \Delta_{\rm{s}}$. Moreover, we observe a modulation of the d$I$/d$V$ intensity with the spin spiral period not only 
outside the superconducting gap, i.e.\ due to the TMR as in Fig.~\ref{fig:Introduction}a and \ref{fig:Introduction}b, but also for the coherence peaks.
The intensity oscillations of the positive and negative energy coherence peaks are in anti-phase, indicating the coherence peaks' opposite spin-polarization.

In Fig.~\ref{fig:Periods}c, we present a line-cut of the d$I$/d$V$ signal averaged over a narrow bias 
voltage
window around zero bias (see the green dashed lines in Fig.~\ref{fig:Periods}b); it reveals the existence of the spin spiral periodicity even deep inside the superconducting gap, exhibiting an approximate cosine dependence (see solid line in Fig.~\ref{fig:Periods}c). 
We interpret this as a manifestation of the spin-polarization of the in-gap states. 

To further investigate the nature of the in-gap electronic states in the studied MSH system, we employ a non-magnetic superconducting tip consisting of a Cr bulk tip with a superconducting Ta-cluster at its apex. Figure~\ref{fig:Periods}e displays the measured d$I$/d$V$ signal as a function of sample bias and position within the spin spiral (see \ref{fig:Periods}d for the expected EMR signal with non-magnetic tip). 
Due to the superconducting tip, the coherence peaks are now shifted to an energy $\pm (\Delta_{\rm{s}}+\Delta_{\rm{t}})$ (outer purple dashed lines; the inner purple dashed lines indicate the tip gap $\pm \Delta_{\rm{t}}$). 
To better visualize the weak spatial variations of the d$I$/d$V$ 
intensity
in Fig.~\ref{fig:Periods}e, 
we average the signal over a small bias range near the positive bias coherence peak (see green dashed lines). The resulting d$I$/d$V$ signal shown in Fig.~\ref{fig:Periods}f reveals a periodicity, which is half of that of the spin spiral.
%
Although the signal variation is much weaker compared to the spin-polarized case,
this result demonstrates that 
the spin spiral
induces an EMR-like contrast with half the magnetic wavelength even for states 
inside the superconducting gap.

\begin{figure}[!htbp]
\centering
\includegraphics[width=\columnwidth]{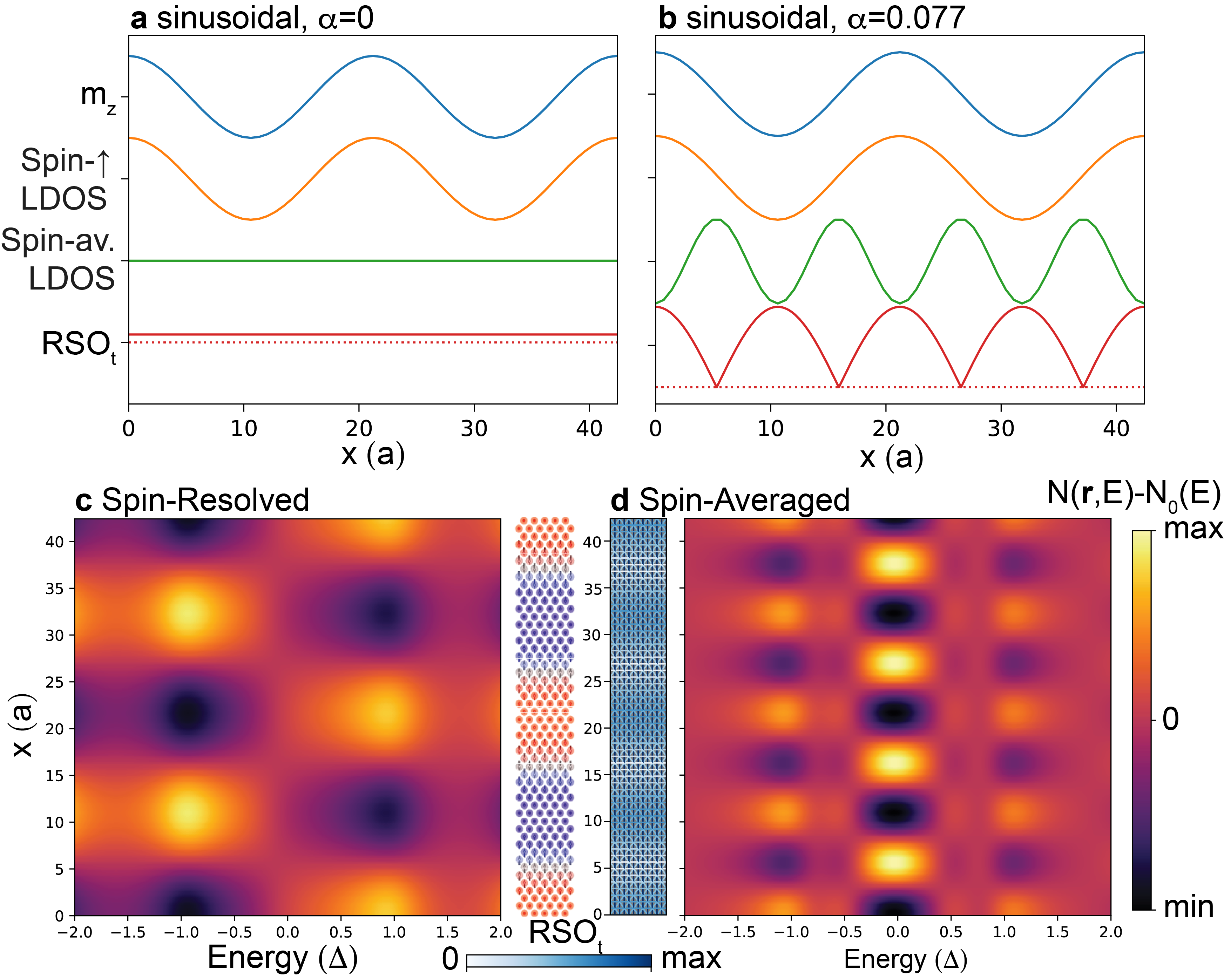}
\vspace{0.01cm}
\captionsetup{font={small}}
\caption{\textbf{Microscopic origin of the TMR and EMR signal.} Line-cut of the out-of-plane magnetization (blue), spin-$\uparrow$ LDOS (orange), spin-averaged LDOS (green) and total Rashba SOC (red) for a sinusoidal spin spiral with \textbf{a} $\alpha = 0$ and \textbf{b} 
$\alpha = 0.077 \Delta$. 
For the case shown in \textbf{b}, the energy- and position-dependent \textbf{c} spin-resolved and \textbf{d} spin-averaged LDOS for a line-cut along the spiral direction of propagation, together with the spatial dependence of $m_z$ and the total Rashba SOC, are shown.}
\label{fig:Theory}
\end{figure}
\noindent

To obtain a better understanding of the physics emerging from the coexistence of spin spiral order and superconductivity, and the microscopic electronic origin of the TMR and EMR signals, we model the MSH system using a minimal Hamiltonian of the form
\begin{align}
    \mathcal{H} =& \; - \mu \sum_{{\bf r}, \alpha} c^\dagger_{{\bf r}, \alpha} c_{{\bf r}, \alpha}
+ \Delta \sum_{{\bf r}} \left(c^\dagger_{{\bf r}, \uparrow} c^\dagger_{{\bf r}, \downarrow} + c_{{\bf r}, \downarrow} c_{{\bf r}, \uparrow}\right) \nonumber \\
&+ \sum_{{\bf r}, {\bf r'}, \alpha} t_{{\bf r}, {\bf r'}} c^\dagger_{{\bf r}, \alpha} c_{{\bf r'}, \alpha}
+ J \sum_{{\bf r}, \alpha, \beta} c^\dagger_{{\bf r}, \alpha} \left({\bf S}({\bf r}) \cdot \boldsymbol{\sigma} \right)_{\alpha,\beta} c_{{\bf r}, \beta} \nonumber \\
&-  \mathrm{i} \sum_{ {\bf r}, {\bf r'}, \alpha, \beta} \alpha_{{\bf r},{\bf r'}} c^\dagger_{{\bf r}, \alpha} \left(\hat{e}_{{\bf r}-{\bf r'}} \times \boldsymbol{\sigma}_{\alpha,\beta} \right)_z c_{{\bf r'}, \beta} \; ,
\end{align}
where $c^\dagger_{{\bf r}, \alpha}$ creates an electron of spin $\alpha$ at site ${\bf r}$, $\mu$ is the chemical potential and $\Delta$ the $s$-wave superconducting order parameter. We denote by $t_{{\bf r}, {\bf r'}}$ the electronic hopping amplitudes with different values for nearest ($t_0$), next-nearest ($t_1$) and next-next-nearest neighbor ($t_2$) sites. $J$ is the magnetic exchange coupling between the spins of the Fe film, and that of the conduction electrons. $\alpha_{{\bf r},{\bf r'}}$ is the 
Rashba spin-orbit (RSO) coupling strength between electrons at sites ${\bf r}$ and ${\bf r'}$, where we consider the same sets of bonds as for the hopping, which arises from the broken inversion symmetry on the surface. 

To identify the microscopic origin of the experimental TMR and EMR signals, we consider two different cases: a homogeneous (sinusoidal) spin spiral a) without RSO (this case had previously been considered in Ref.\cite{NakosaiPRB2013}), and b) with a non-zero RSO. 
In Figs.~\ref{fig:Theory}a and \ref{fig:Theory}b, we present the corresponding out-of-plane magnetization, $m_z$, of the spin spiral (blue), together with the resulting zero-energy spin-resolved (orange) and spin-averaged (green) LDOS
for these two cases along a line-cut parallel to the spiral propagation direction ${\bf Q}$. We note that the spin-resolved and spin-averaged LDOS are directly related to the experimentally obtained d$I$/d$V$ signals with a spin-polarized (TMR) and a non-magnetic (EMR) tip, respectively. Moreover, the non-collinear nature of the spin structure induces an additional Rashba spin-orbit interaction, RSO$_i$, such that topological superconductivity can arise even in the absence of a conventional RSO~\cite{NakosaiPRB2013,BedowPRB2020,MascotQ-MAT2021}. The induced RSO$_i$ in combination with the conventional RSO leads to a total RSO$_t$, shown as a red line in Figs.~\ref{fig:Theory}a and \ref{fig:Theory}b, that determines the emerging properties of the system.

For both cases, the spin-resolved LDOS follows the spatial form of $m_z$, thus capturing the experimentally observed spin spiral period observed via TMR (see Fig.~\ref{fig:Periods}c). In contrast, the spin-averaged LDOS reflects the spatial form of RSO$_t$. This leads to a spatially constant LDOS for case a), in disagreement with the experimentally observed EMR signal (Fig.~\ref{fig:Periods}f). However, for case b), the spin-averaged LDOS exhibits a spatial modulation with a periodicity of half of the spin spiral, in agreement with the experimental EMR results. We note that for the case of Fig.~\ref{fig:Theory}b, the complex interplay between the conventional and the induced RSO leads to this spatially modulated RSO$_t$. We thus attribute the origin of the experimentally observed EMR signal to the interplay between the RSO coupling and the spin spiral. While we explore the consequences of this interplay further below, we note that an alternative origin of the EMR signal could be given by the presence of an inhomogeneous spin spiral even in the absence of an RSO coupling.

In Figs.~\ref{fig:Theory}c and \ref{fig:Theory}d, we present a line-cut of the energy-dependent spin-resolved and spin-averaged LDOS, respectively, along ${\bf Q}$ (to highlight the spatial oscillation for both cases, we have subtracted the respective spatially averaged LDOS, N$_0$(E)). In the spin-resolved case, we find that the LDOS at the position of the coherence peaks is modulated with the spin spiral period, with an enhancement of the LDOS at one coherence peak accompanied by a suppression at the other. This complementary pattern agrees well with the oscillatory pattern of the TMR signal at $\pm \Delta_s$ shown in Fig.~\ref{fig:Periods}b, and reflects the partial, opposite spin polarization of the coherence peaks.
The spin-averaged LDOS, Fig.~\ref{fig:Theory}d, also shows a spatially oscillating pattern, however, with a period that is half of that of the spin spiral, as observed experimentally with a non-magnetic tip (Fig.~\ref{fig:Periods}f). In addition, these results exhibit a complementary intensity pattern between the coherence peaks and the low energy region around $E_F$, with the region of largest RSO$_t$ interaction coinciding with the lowest LDOS at $E_F$ (see left side of the image in Fig.~\ref{fig:Theory}d). 
These results suggest, that the experimentally accessible spin-averaged LDOS in the superconducting gap provides direct access to the complex spatially-dependent total RSO$_t$, a parameter that governs the emerging properties of non-collinear MSH systems.

\begin{figure*}[!htbp]
\centering
\includegraphics[width=0.9\textwidth]{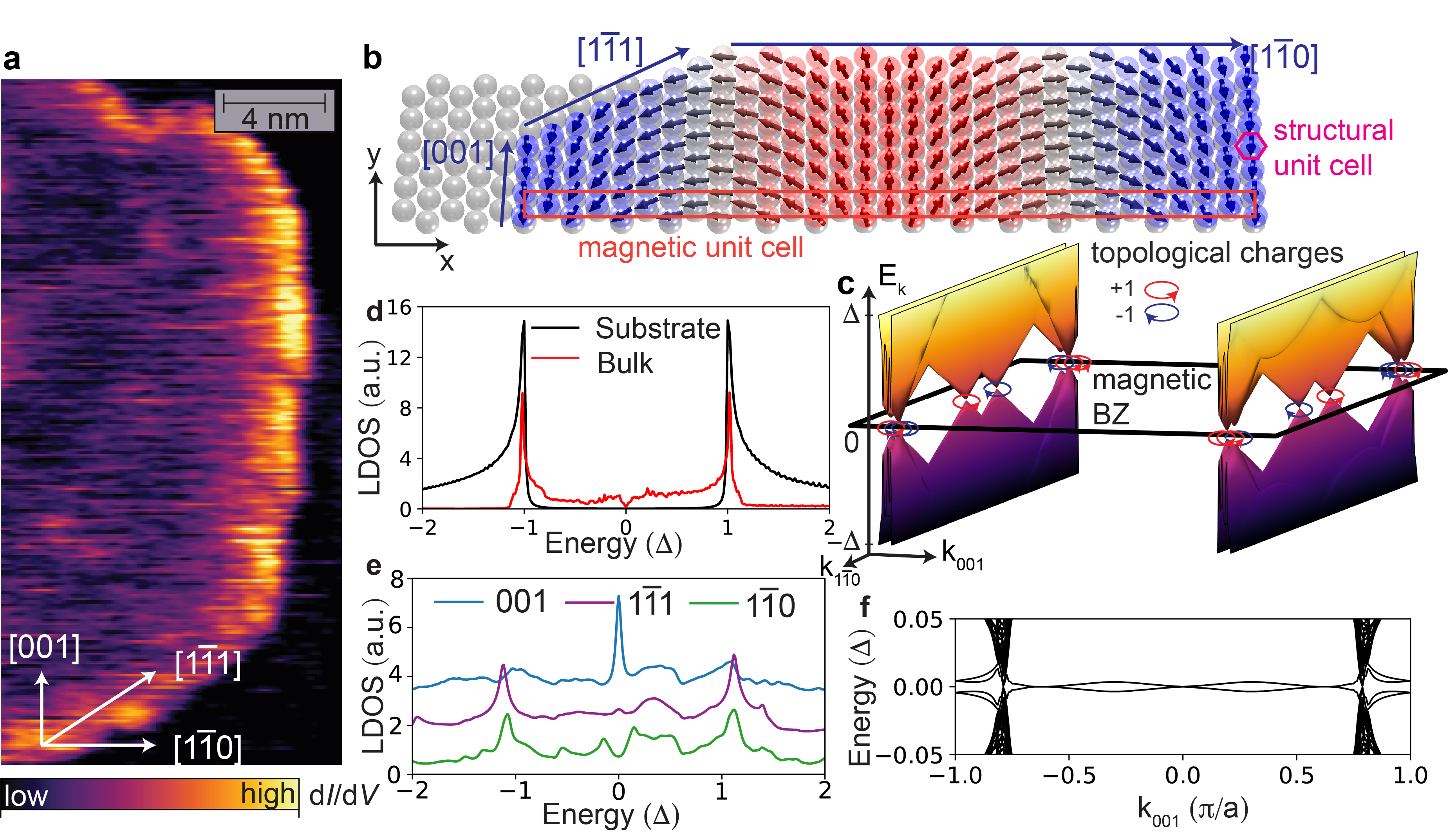}
\vspace{0.2cm}
\captionsetup{width=\linewidth, font={small}}
\flushleft
\caption{
\textbf{Topological nodal point superconductivity and edge modes in the MSH system.}
\textbf{a} Spin-averaged in-gap d$I$/d$V$ map at $V=0.1$~mV exhibiting an enhanced contrast along the [001]-edge.
\textbf{b} Sketch of the structural and magnetic unit cell. \textbf{c}  Electronic structure in the magnetic Brillouin zone exhibiting nodal points with non-zero topological charge. \textbf{d} Theoretical LDOS on the Ta surface (black) and for the Fe/Ta MSH system (red). \textbf{e} Theoretical LDOS at a [001]-edge (blue), [1$\bar{1}$1]-edge (purple) or [1$\bar{1}$0]-edge (green). \textbf{f} Electronic band structure as a function of momentum along the [001]-edges of a ribbon system. The spin spiral terminates with an angle $\theta = 0^{\circ}$ at the ribbon's left and right edges.}
\label{fig:Theory2}
\end{figure*}

\noindent

Next, we explore the spectroscopic signatures along the edges of the Fe islands, as shown in Fig.~\ref{fig:Theory2}a, where we present a d$I$/d$V$ map obtained at $V=0.1$~mV for one of the investigated Fe islands. The observed strongly enhanced intensity in particular at the [001] edge raises the intriguing question of whether it is related to the existence of a topological edge mode, as has previously been found in collinear MSH systems~\cite{BazarnikNAT-COMM2023}. To investigate this question, we plot in Fig.~\ref{fig:Theory2}c the calculated electronic band structure of the system, which exhibits several nodal points in the magnetic Brillouin zone (the corresponding magnetic unit cell in real space is shown in Fig.~\ref{fig:Theory2}b).
As these nodal points possess a quantized topological charge, $q=\pm 1$, we can conclude that the system is in a topological nodal point superconducting phase. As a result, the LDOS of the bulk Fe/Ta system exhibits a characteristic $V$-shape around zero energy, directly reflecting the existence of these nodal points, in contrast to the uncovered Ta surface, where the LDOS reflects the existence of a hard $s$-wave superconducting gap (see Fig.~\ref{fig:Theory2}d). 

A unique feature of the TNPSC phase is that it exhibits zero energy modes only along certain real space edges, which are determined by the interplay of the nodal point position in momentum space, their projection onto the edge direction, and the magnetic structure of the edge \cite{BazarnikNAT-COMM2023,Zhang2019,KieuPRB2023}. Fe/Ta(110) islands in general realize three different types of edges along the $[001]$, $[1{\bar 1}1]$, and $[1{\bar 1}0]$ directions, as shown in Fig.~\ref{fig:Theory2}b. A comparison of the LDOS for these three edges (see Fig.~\ref{fig:Theory2}e) reveals that a 
zero-energy peak occurs only along the [001] edge, but is absent for the other two edges. This peak arises from a low-energy, weakly dispersing chiral edge mode that connects nodal points of opposite topological charge, as can be seen in the plot of the electronic band structure of a ribbon system with [001] edges shown in Fig.~\ref{fig:Theory2}f. This finding 
suggests that the enhanced d$I$/d$V$ along the edges of the Fe islands shown
in Fig.~\ref{fig:Theory2}a reflects the existence of a chiral edge mode. 

\begin{figure*}[!htbp]
\centering
\includegraphics[width=0.9\textwidth]{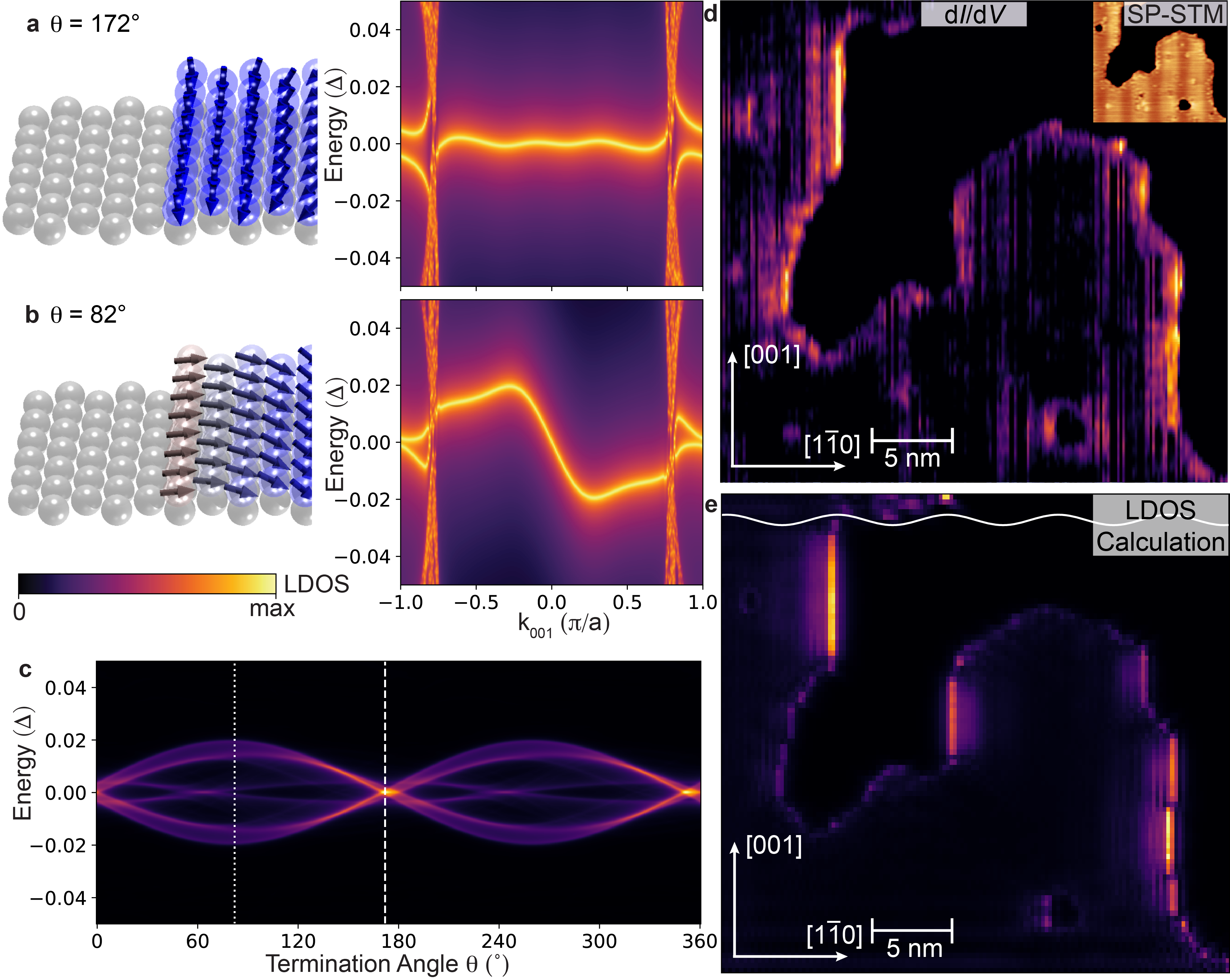}
\vspace{0.2cm}
\captionsetup{font={small}}
\caption{\textbf{Edge states along [001] direction.}
Spectral function as a function of momentum along a ribbon with two [001]-edges and termination angle \textbf{a} $\theta =172^{\circ}$ and \textbf{b} $\theta =82^{\circ}$ of the spin spiral. \textbf{c} Low-energy LDOS at an [001]-edge as a function of the termination angle $\theta$. The white dashed and dotted lines represent the $\theta =172^{\circ}$ and $\theta =82^{\circ}$ terminations, respectively.
\textbf{d} Experimental spin-averaged zero-bias d$I$/d$V$ map with fast scan direction along [001]-edges of an irregularly shaped Fe island; the inset shows an SP-STM constant-current image of the same area. 
\textbf{e} Theoretical spin-averaged zero-energy LDOS for a Fe island of the same size and shape as that shown in \textbf{d}. The white line represents the $m_z$-component of the spin spiral along the $[1{\bar 1}0]$-direction.
}
\label{fig:Edge_States}
\end{figure*}

Until now, chiral edge modes have been observed only in collinear MSH systems. In our non-collinear case, the termination of the cycloidal spin spiral by a [001] edge  
can lead to an arbitrary angle $\theta$ between the spin direction at the edge and the surface normal, as schematically shown for two values of $\theta$ in Figs.~\ref{fig:Edge_States}a and \ref{fig:Edge_States}b. 
This naturally raises the question of whether the termination angle $\theta$ bears any effect on the properties of the edge mode. To investigate this question, 
we present in Figs.~\ref{fig:Edge_States}a and \ref{fig:Edge_States}b the momentum- and energy-resolved spectral function at a ribbon's left edge for two different termination angles of $\theta = 172^\circ$ and $\theta = 82^\circ$, respectively. These results demonstrate that the edge mode dispersion can indeed be changed by varying $\theta$, with the least and most dispersive edge modes obtained for the cases shown in Figs.~\ref{fig:Edge_States}a and \ref{fig:Edge_States}b. 

This change in the dispersion with varying $\theta$ is directly reflected in the form of the low-energy LDOS, as shown in Fig.~\ref{fig:Edge_States}c, where we present the energy-resolved LDOS at a [001] edge as a function of $\theta$. While for $\theta = 172^\circ$, the LDOS exhibits a strong peak at zero-energy, corresponding to the very weakly dispersing edge mode shown in Fig.~\ref{fig:Theory2}f, the more dispersive edge mode for $\theta = 82^\circ$ results in two peaks in the edge LDOS at non-zero energy, and a much weaker LDOS at zero energy. This termination-dependent edge mode is a unique feature of the spiral magnetic structure, not seen before in any of the MSH systems possessing collinear magnetic ground states. This feature could in principle be used to reverse the chirality of an edge mode, by sliding the spiral across the systems such that the termination angle changes by 180$^\circ$. 

\noindent

These intriguing results -- the existence of a low-energy chiral edge mode at a [001] edge, and the absence thereof at $[1{\bar 1}1]$ and $[1{\bar 1}0]$ edges, as well as a variation in intensity of the zero-energy LDOS at a [001] edge with the termination angle -- are unique and experimentally observable features of this TNPSC phase. To test these theoretical predictions, we study an Fe island (whose SP-STM image is shown in the inset of Fig.~\ref{fig:Edge_States}d) that possesses several [001] and $[1{\bar 1}1]$ edges. A measurement of the zero-bias d\textit{I}/d\textit{V}, shown in the main panel of Fig.~\ref{fig:Edge_States}d, reveals two main results: a large intensity along the [001]-edges, which varies between different [001]-edges, and a very weak intensity along the [1$\bar{1}$1]-edges. These findings provide strong evidence for the existence of a TNPSC phase, and for the predicted termination angle dependent zero-energy LDOS at [001] edges shown in Fig.~\ref{fig:Edge_States}c. To directly compare our experimental findings with our theoretical model
, we compute the zero-energy LDOS for an Fe island of the same size and shape as the experimental one, as shown in Fig.~\ref{fig:Edge_States}e. A comparison of the experimental d$I$/d$V$ in Fig.~\ref{fig:Edge_States}d with the theoretical LDOS in Fig.~\ref{fig:Edge_States}e shows very good agreement, both in terms of a large intensity along [001]-edges only, as well as a variation in intensity along different [001]-edges, providing further support for the existence of a TNPSC phase in Fe/Ta(110).

In conclusion, our combined experimental and theoretical study of the spin spiral state of an Fe monolayer proximity-coupled to a superconducting Ta(110) substrate revealed an enhanced LDOS at [001]-edges of Fe islands which can be attributed to a chiral edge mode resulting from topological nodal-point superconductivity. Due to the non-collinearity of our spin texture, different magnetization directions can occur at the structurally identical [001] edges, which strongly affects the dispersion of the chiral edge mode. We expect that our finding of magnetization-direction dependent chiral edge modes can be used as an additional knob for future application concepts in the realm of topological superconductivity and superconducting spintronics.

\bmhead{Acknowledgments}
R.B. and R.W. acknowledge funding by the European Research Concil (ERC Advanced Grant No. 786020).
K.v.B.\ acknowledges financial support from the Deutsche Forschungsgemeinschaft (DFG, German Research Foundation) via project no. 402843438. J.B.\ and D.K.M.\ acknowledge support by the U.\ S.\ Department of Energy, Office of Science, Basic Energy Sciences, under Award No.\ DE-FG02-05ER46225.
J.B., D.K.M. acknowledge financial support for guest stays in Hamburg by the DFG via the Cluster of Excellence "Advanced Imaging of Matter" (EXC 2056-project ID 390715994). R.W. acknowledges funding by the DFG via the Cluster of Excellence "Advanced Imaging of Matter" (EXC 2056-project ID 390715994). R.L.C.\ acknowledges financial support from the Deutsche Forschungsgemeinschaft (DFG, German Research Foundation) via grant no. 459025680.

\bibliography{sn-bibliography}
\newpage
\end{document}